\documentstyle[preprint,aps,pre,eqsecnum]{revtex}     
\tighten
\input epsf
\begin{document}
\draft

\title{Universal Short-Time Dynamics in the\\ Kosterlitz-Thouless Phase}
\author{P. Czerner and U. Ritschel}

\address{Fachbereich Physik, Universit\"at GH Essen, 45117 Essen
(F\ R\ Germany)}
\date{submitted to Phys. Rev. E}

\maketitle
\narrowtext

\begin{abstract}
We study the short-time dynamics of
systems that develop ``quasi long-range order''
after a quench to
the Kosterlitz-Thouless phase. With the working hypothesis
that the ``universal short-time behavior'', previously
found in Ising-like systems, also occurs in
the Kosterlitz-Thouless phase, we explore the scaling behavior
of thermodynamic variables during the relaxational process
following the quench.
As a concrete example,
we investigate the two-dimensional
$6$-state clock model by Monte Carlo simulation.
The exponents governing the magnetization,
the second moment, and the autocorrelation function are calculated.
{}From them, by means of scaling relations, estimates for
the equilibrium exponents $z$ and $\eta$ are derived.
In particular, our estimates
for the temperature-dependent
anomalous dimension $\eta$ that governs the static
correlation function are consistent with existing analytical
and numerical results and, thus,
confirm our working hypothesis.
\end{abstract}
\pacs{PACS: 64.60.Ht,75.40.Gb,67.40.Fd,02.70.Lq}
\section{Introduction}
As discovered by Janssen et al. \cite{jans},
universal short time
behavior (USTB) in critical dynamics
occurs after quenching a spin system
with nonconserved order parameter (model A \cite{halp}) from
a high-temperature initial state to the
its critical point. In its most pronounced form the USTB manifests itself
in the time dependence of the order-parameter expectation value.
Starting from a small average
initial magnetization $m_0$, the order starts to grow during
a nonequilibrium (though universal) period shortly
after the quench, before the decay to the
equilibrium value zero sets in. The initial increase is governed
by a power law,
\begin{mathletters}\label{order}
\begin{equation}\label{power}
m\sim m_0\,t^{\theta}\>,
\end{equation}
and the short-time exponent $\theta$ is given by \cite{own,foo1}
      \begin{equation}\label{theta}
\theta=\frac{\eta_0-\eta}{2z}\>,
\end{equation}
\end{mathletters}
i.e., it is determined
by the difference between the anomalous dimensions
$\eta_0$ and $\eta$ of the
{\it initial} order-parameter field and the equilibrium magnetization,
respectively. ($z$ denotes the dynamic equilibrium exponent.)
$\theta$ (or, equivalently $\eta_0$) is a new exponent, not
expressible in terms of known static and/or dynamic exponents \cite{jans}.
Thus, the initial increase will only occur under the condition that
$\eta_0>\eta$, which, for instance, is fulfilled in the case of
the Ising model for spatial dimensions $1 < d <4$.

Recently significant progress has been made in quantitative studies
of the USTB by means of
Monte Carlo (MC)
simulations.
It turned out to be possible to observe the short-time power law
(\ref{power}) in
surprisingly small systems and after
relatively short simulation times\cite{monte,gras}.
As the exponents governing the short-time behavior of
thermal averages can be expressed in terms of scaling
relations among short-time and equilibrium exponents,
it is possible to extract even static
exponents from a relatively early stage of the dynamics,
whereby the problem of critical
slowing down is circumvented to some extent\cite{liea,sczh}.

As pointed out by Janssen \cite{jans2},
the qualitative mechanism underlying the USTB is
the following: After the quench
the initially small correlation length
starts to grow towards its equilibrium value.
During the early stages, when
fluctuations are still short-ranged, a mean-field
description of the process should hold. Since in Ising-like
systems
$T_c^{\text{MF}}$ is always larger than the real $T_c$,
it is plausible that shortly after the quench a spontaneous
magnetization will evolve, pointing in the direction of the
initial magnetization $m_0$. At some point, when $\xi(t)$ has
grown to a certain size, this increase will come to a halt, and
afterwards the
decay to the equilibrium value $m(t\to \infty)=0$ will begin.

So far in both analytical and numerical calculations the USTB has only
been studied in Ising-like systems, with a
conventional critical (or tricritical)
point\cite{jans,own,monte,gras,liea,sczh,jans2,oerding,tricrit}.
However, the phenomenon
(and the applications built thereon) should not only
occur at a
critical point, but, more generally, in situations
where scale-invariant long-ranged fluctuations evolve
after a quench
from a high-temperature initial state
(or after any other comparable change of the state of the system).
In the following this scenario is tested for
a quench to the Kosterlitz-Thouless (KT) phase?
In two-dimensional systems with $O(2)$ symmetry (or variants
of it) below the KT transition temperature $T_{\text{KT}}$,
one finds a range of temperatures with a
power-law decay of the static
correlation function\cite{koth,jose}. Examples are the $XY$ model
and the $p$-state
clock model for $p\ge 5$.
Additionally, mean-field theory predicts
for these systems (erroneously) a conventional
critical point, and one finds generally $T_c^{\text{MF}}>T_{\text{KT}}$.
Thus, taking into account the foregoing discussion, the USTB
should also be observable after a quench
to the KT phase, and it is interesting to see whether the methods
described in Refs. \cite{monte,gras,liea,sczh} can be applied, i.e., whether
for instance the equilibrium exponent $\eta$ can be extracted from
the short-time relaxational behavior.

The rest of this paper is organized as follows.
In the next Section
we recall analytical and numerical results for the clock model.
In Sect.\,III
we derive the scaling behavior of the relevant thermodynamic
quantities, taking into account the anomalous scaling of the
initial magnetization. Our own simulations have been
carried out for the $p=6$ clock model.
The numerical results are presented in Sect.\,IV.

\section{The clock model}

According to the Mermin-Wagner theorem \cite{mewa}, the
two-dimensional {\it XY} model cannot undergo a conventional
para-/ferromagnetic phase transition. However, as shown by
Kosterlitz and Thouless \cite{koth}, there exists a transition
between a state with unbounded and bounded vortices
at $T=T_{\text{KT}}$.

A model that in many respect interpolates between the Ising model
and the {\it XY} model is the $p$-state clock model, defined by the
Hamiltonian
\begin{equation}\label{clock}
H=-J\,\sum_{<ij>}\;\vec s_i\cdot\vec s_j=
-J\sum_{<ij>}\;\cos\left(\vartheta_i-\vartheta_j\right)\,,\quad
\vartheta_i=\frac{2\pi}{p}\,n_i\;,
\end{equation}
where $n_i\in \left\{0,1,\ldots,p-1\right\}$
and $<.>$ denotes the sum over nearest
neighbors. This spin system is equivalent
to the {\it XY} model for $p\to \infty$ and to the Ising model
for $p=2$. It was first shown by Jose et al.
\cite{jose} and later on verified by numerous MC
studies \cite{p6cl,tobo,kagu,chal,yaon,lero}
that the $p$-state clock model has
a single Ising-like critical point for $p\le 4$.
For $p\ge 5$, on the other hand,
there exist two transition points, with temperatures $T_1$
and $T_2(>T_1)$, respectively. For $T>T_2$ a paramagnetic
high temperature phase exists, and for
$T<T_1$ the system becomes ferromagnetic.
For $T_1<T<T_2$, however,
a KT phase with ``quasi long-range order'' emerges, where
the correlation function behaves as
$C({\bf r})\sim |{\bf r}|^{-\eta}$, with the
anomalous dimension $\eta=\eta(T)$ continuously depending
on the temperature \cite{jose}.
Both transitions are of KT type, i.e.,
the correlation length diverges exponentially as $T_1$ $(T_2)$
is approached from below (above)\cite{jose}.

Our MC simulations---the results are presented in
Sect.\,IV---were done for the $p=6$-state clock model. The case $p=6$
was studied in great detail in the literature, partly motivated
by the fact that
in real two-dimensional systems one expects
anisotropies  with six-fold symmetry due to the lattice
structure\cite{jose}.
Approximate analytic results for
the transition temperatures and the exponent $\eta(T)$ have
been obtained by Jose et al. \cite{jose}. While $T_2$ was found
to be identical with
the KT transition temperature of the $XY$ model, T/J=0.95,
$T_1$ was predicted to vary as $T_1/J=4\pi^2/1.7p^2$ (which
means $T_1/J=0.645$ for $p=6$).
(Here and in the following we set the Boltzmann constant $k_B=1$.)
The exponent
$\eta$ was found to vary between $\eta(T_2)=1/4$ and
$\eta(T_1)=4/p^2$ (which means $\eta(T_1)=0.111$ for $p=6$).

Among the MC analyses mentioned above, we used the
results of Challa and Landau \cite{chal}
for comparison with our own data. They found for the
transition temperatures $T_2/J=0.92(1)$ and $T_1/J=0.68(2)$ and
for the corresponding exponents
$\eta(T_2)=0.275$ and $\eta(T_1)=0.100$.
(No errors for $\eta(T)$ are given in Ref.\,\cite{chal}.)
Both the results for the transition temperatures and the exponents
are consistent with the analytic results of Ref.\,\cite{jose}.

Dynamic MC simulations for the 6-state clock model
were performed by Kaski et al. \cite{kagu}, but in this
work the relaxational behavior in the ferromagnetic phase was studied,
with emphasis on the growth of the domain size after a quench.
Thus, concerning the temperature ranges,
this work is complementary to ours.

In order to comply with the requirements
for the occurrence of the USTB
discussed in the Introduction, the system has to have a mean-field
$T_c^{\text{MF}}$ above the temperature to which the quench is
carried out. In the case of the clock model it is straightforward
to show that there exists a $T_c^{\text{MF}}$ for all $p\ge 2$
and that $T_c^{\text{MF}}/J= 2 d$ for $p=2$ (Ising case)
and $T_c^{\text{MF}}/J= d$ for $ p > 2$, where
$d$ denotes the spatial dimension. Comparison with
exact or numerical results for the clock models shows
that the upper transition temperature
(which is a critical point for $p=2,\,3,\,4$) is below
$T_c^{\text{MF}}$ for all $p\ge 2$.
E.g. for $p=6$, the upper transition temperature $ T_2/J \simeq 0.92$
is well below $T_c^{\text{MF}}/J=2$.

\section{Scaling Analysis}

In the KT phase, i.e. for all temperatures below $T_{\text{KT}}$,
the bulk {\it equilibrium} correlation length is formally infinite.
After a sudden quench from a high-temperature initial state to
the KT phase, the initially small correlation length $\xi(t)$
starts to grow, and the large-scale fluctuations gradually
develop.
Let us first
discuss relaxational processes in Ising-like systems.
In this case,
after a short period of nonuniversal behavior following the quench,
the growth of
the correlation length is described by the
universal power law,
\begin{equation}\label{xioft}
\xi(t)\sim t^{1/z}\>,
\end{equation}
and in the bulk system $\xi(t)$ becomes ``macroscopic''
for large times.
Moreover, in the universal regime $\xi(t)$ is the
{\it only} macroscopic time scale and allows to
characterize the time dependence of correlation functions completely.
In the finite systems, (\ref{xioft}) holds
as long as $\xi(t) <L$ (where $L$
is the linear system dimension). In the following
this scenario will be called
simple (dynamic) scaling.

In the KT phase
vortices are generated in the course of the quench
and potentially disturb
the simple scaling picture\cite{Mondello,Yurke,blun}.
Deviations from the simple scaling picture
are expected mainly for intermediate times.
As we shall see from the evaluation
of the MC data, modifications compared to the Ising dynamics
really occur in certain quantities in form of a relatively slow
approach to the power-law form. Below we review the simple
scaling scenario concerning the quantities
studied in our MC simulation, with emphasis on the
(expected) consequences of the USTB. Modifications due to vortices
are discussed in some more
detail in the context of the MC simulation in Sect.\,IV.

\subsection{Bulk behavior}

Consider first the expectation value of the magnetization density\cite{foo3}.
Suppose in the high-temperature initial state an external magnetic field
has generated a small (initial) magnetization $m_0$, pointing in the
$\vartheta=0$ direction of the $xy$ plane.
Then, after the quench, the expectation value of the order
parameter, defined by
\begin{equation}\label{orderdef}
m(t) := L^{-2}\langle \sum_{i} \cos\vartheta_i\rangle\>,
\end{equation}
should remain in the direction of the initial magnetization.
In the scaling regime it should further satisfy
the homogenity property\cite{foo5}
\begin{equation}\label{ordersc}
m(t,\,m_0)\approx l^{\eta/2}\;m(l^zt,\,l^{-\eta_0/2}m_0)\>,
\end{equation}
where $l$ is a rescaling parameter, and the exponents $\eta,\,\eta_0$
were introduced in Eqn.\,(\ref{power}).

{}From Eqn. (\ref{ordersc}) it follows that, asymptotically,
the magnetization behaves as
\begin{equation}\label{asymp}
m(t,\,m_0)\sim t^{-\eta/2z}\,{\cal F}(m_0\,t^{\eta_0/2z})\;,
\end{equation}
where ${\cal F}(x)$ is a universal scaling function. As an immediate
consequence of $\eta_0\neq\eta$ in Eqn. (\ref{ordersc}), the
rescaled (or renormalized)
$m_0$ is no longer the initial value of $m(t)$. However,
one  still has asymptotically $m\propto m_0$
for $t\to 0$\cite{jans,jans2}. Thus the scaling
function ${\cal F}(x)$ in (\ref{asymp}) behaves $\sim x$ for
$x\to 0$, and, as a result, the short-time behavior is described
by (\ref{order}).

Further, from the combination in the argument of ${\cal F}$ in
(\ref{asymp}) one realizes that $m_0$ also gives rise to a novel
time scale, $t_0\sim m_0^{-2z/\eta_0}$. While for $t\lesssim t_0$
the USTB (\ref{power}) is observed, the system crosses over to
the nonlinear decay $m\sim t^{-\eta/2z}$ for $t\gtrsim t_0$, which
implies that ${\cal F}(x)\to \mbox{const.}$ for $x\to \infty$ and
that memory of the initial value $m_0$ is lost for $t\gg t_0$.

Can we expect to observe the USTB even for $m_0=0$?
Obviously, in this case the magnetization vanishes identically
for $t>0$. However, as demonstrated in Ref.\,\cite{jans2},
the USTB leaves its fingerprint on the relaxational
behavior of the autocorrelation function
\begin{equation}\label{auto}
A(t)=\langle \vec s_i(t)\cdot \vec s_i(0)\rangle\>,
\end{equation}
which measures the change the system has undergone during the time
$t$ after the quench. Without going into the technical details, we
here just present Janssen's result \cite{jans2}:
\begin{equation}\label{lambda}
A(t)\sim t^{-\lambda/z}\>,\qquad {\text{with}}\quad \lambda=2-z\theta\>,
\end{equation}
i.e., the short-time exponent $\theta$ also appears in $A(t)$
and leads to a deviation from the naive expectation $\lambda=d$,
surprisingly for all $t>0$. It
was this quantity where the USTB actually was first observed in a
MC simulation \cite{huse}.

The third quantity that will play a role in the following is the
second moment (or variance) of the magnetization. It is defined by
\begin{equation}\label{mtwo}
m^{(2)}(t)=L^{-2}\langle\>\sum_{ij}\!\vec s_i(t)\cdot\vec s_j(t)\;\rangle\>.
\end{equation}
A straightforward analysis reveals that $m^{(2)}(t)$, again for
vanishing initial magnetization, takes the scaling form
\begin{equation}\label{b}
m^{(2)}(t)\sim t^{\,b}\>,\qquad{\text{with}}\quad b=\frac{2-\eta}{z}\>,
\end{equation}
i.e., the USTB does not influence this quantity\cite{own3}.

As suggested by Sch\"ulke and Zheng\cite{sczh}, the three exponents
$\theta$,
$\lambda$, and $b$, governing the bulk behavior of $m$, $A$, and
$m^{(2)}$, respectively,
suffice to determine the {\it equilibrium} quantities $\eta$
and $z$ from the USTB. The starting point is the short-time
exponent itself, which can be obtained from simulations
with small, nonvanishing $m_0$. Then, from simulations with
$m_0=0$, the exponents $\lambda$ and $b$ can be estimated
simultaneously. Utilizing the scaling relation
(\ref{lambda}), $z$ can be determined, and, in turn, with
relation (\ref{b}) eventually the scaling dimension $\eta$.

Our main hypothesis for the present work is that also in the KT phase,
analogously to the situation in the critical
Ising model, the operator
dimension of the initial field, $\eta_0$, is different
from $\eta$, and that, as a consequence, the scaling picture
developed above also describes the short-time dynamics in the
KT phase.
This is neither obvious nor by any means rigorously
proved. At the present stage it is merely a working hypothesis,
which below will be tested by means of MC
simulations.
Especially for quenches to the KT phase, we expect all exponents
to depend on the respective final temperature to which the quench
is carried out.

\subsection{Finite-size scaling}

Starting from Eqn. (\ref{ordersc}), all the results about the
scaling behavior of the thermodynamic quantities described
so far hold in the bulk system. If we are dealing with a system of
finite size---and certainly this is the case in the MC
simulation---the power laws (\ref{power}), (\ref{lambda}),
and (\ref{b}) hold as long as the time-dependent correlation
length $\xi(t)\sim t^{1/z}$ is smaller than the system size $L$.
A finite-size scaling analysis reveals how
the thermodynamic quantities behave during the later stages of
the relaxational process\cite{own,own2}.

Taking into account the finite system size, the analog to (\ref{ordersc})
reads
\begin{equation}
m(t,m_0,L)\approx l^{\eta/2}m(l^z t,l^{-\eta_0/2}m_0,lL)\>.
\end{equation}
Removing the rescaling parameter $l$ by setting $l\sim L$, one obtains
\begin{equation}\label{asymfs}
m(t,m_0,L)\sim L^{-\eta/2}\;{\cal G}(t/L^z,m_0L^{\eta_0/2})\>,
\end{equation}
where ${\cal G}(x,y)$ is a universal scaling function. As seen
from (\ref{asymfs}),
the size $L$ of the system gives rise to another characteristic
time scale $t_L\sim L^z$ (besides the initial time scale $t_0$),
the well-known finite-size scale\cite{sancho}.
In the KT phase it turns out temperature dependent.
Further, it was shown in
Ref. \cite{own} that ${\cal G}(x,y)\sim {\cal A}(y){\rm e}^{-x}$ for
$x\to \infty$, i.e., the magnetization decays linearly when
$t\gg L^z$. The scaling function ${\cal A}(y)$ in
(\ref{asymfs}) describes the universal dependence of late stages
of the relaxational process on the initial magnetization $m_0$.
This dependence is special to the finite-size system\cite{own2}.In the
bulk memory of the initial condition is lost for $t\to \infty$.

For arbitrary $x$ and $y\to 0$, the function ${\cal G}(x,y)$
behaves as ${\cal G}\sim y\,{\cal B}(x)$, with a scaling function
${\cal B}$. In this limit
the maximum of the magnetization profile has the scaling form
\begin{equation}\label{max}
m_{\text{max}}\sim L^{(\eta_0-\eta)/2}\,m_0\,{\cal B}(t_{\text{max}}/
L^z)\>.
\end{equation}
Thus, the finite-size scaling behavior of $m_{\text{max}}$ allows
in principle
to determine {\it directly} the difference $\eta_0-\eta$.

Also for the other quantities discussed in Sect.\,III.a the asymptotic
behavior is modified when $\xi(t)\gtrsim L$. Firstly, the autocorrelation
function (\ref{auto}) also decays linearly for
$t\gtrsim t_L$. Secondly, the second moment takes the scaling form
\begin{equation}
m^{(2)}(t,L)\sim t^{\, b}\,{\cal J}(t/t_L).
\end{equation}
In order to reproduce (\ref{b}), the scaling function ${\cal J}(x)$ has
to approach a constant for $x\to 0$. For $x\to \infty$, ${\cal J}(x)\sim
x^{-b}$, such that in the long-time limit the equilibrium
result $m^{(2)}\sim L^{2-\eta}$ is obtained.

Again, in the KT phase we expect to observe the finite-size
scaling scenario presented above, with the
characteristic scale $t_L$
depending on the temperature.

\section{Monte Carlo Analyses}
\subsection{The method}
We used the usual time-dependent interpretation of the sampling
algorithm \cite{bind},
with sequential updating and nonconserved total
spin. We employed the heat-bath algorithm, as described in detail
for the clock model in Ref. \cite{lero}. It is equivalent to Glauber
dynamics \cite{bind}. All simulations were carried out on square
lattices with varying linear dimension $L$ with periodic
boundary conditions.

As the initial configurations
we used equilibrium states for uncoupled spins in an external
magnetic field $h=2m_0$ (with
$m_0=0$ for the autocorrelation function and the second moment).
Time-dependent
thermal averages were approximated by summing
over a number of $N$ histories, $N$ varying between 1000 and 10000
depending on the size $L$. In order to obtain a measure
for the error, the $N$ histories were divided in 10-20
runs, from which statistically independent exponent
``measurements'', their mean value, and the statistical error
of the mean value were obtained.

As mentioned in the Introduction,
an attractive feature of the USTB in Ising-like systems
is that relatively relatively small
lattices and short simulation times turned out to be
sufficient extract reliable
results for exponents\cite{monte,liea,sczh}. However,
the estimates obtained will in general to some extent depend
upon the lattice size $L$ and, in the case of $\theta$, upon
the initial magnetization $m_0$. In the case of the clock
model, it turned out that one has to go to somewhat larger
lattices and, especially in the case of the autocorrelation function,
the time until the power-law behavior is assumed is increased compared
to the Ising or Potts model.

\subsection{The short-time exponent $\theta$ and the linear relaxation time}

Firstly we discuss our data for the magnetization and the determination
of the short-time exponent $\theta$ from them. In order
to get an impression of the time dependence of the
order parameter, we have calculated relaxation
profiles of the 6-state clock model
for $L=30$ and $m_0=0.05$
in the temperature range $0.5\le T/J\le 1.1$, i.e., the lower
limit lies well below $T_1$
in the ferromagnetic phase and the upper limit
above $T_2$ in the paramagnetic phase.
\\[3mm]

\def\epsfsize#1#2{0.58#1}
\hspace*{-0.4cm}\epsfbox{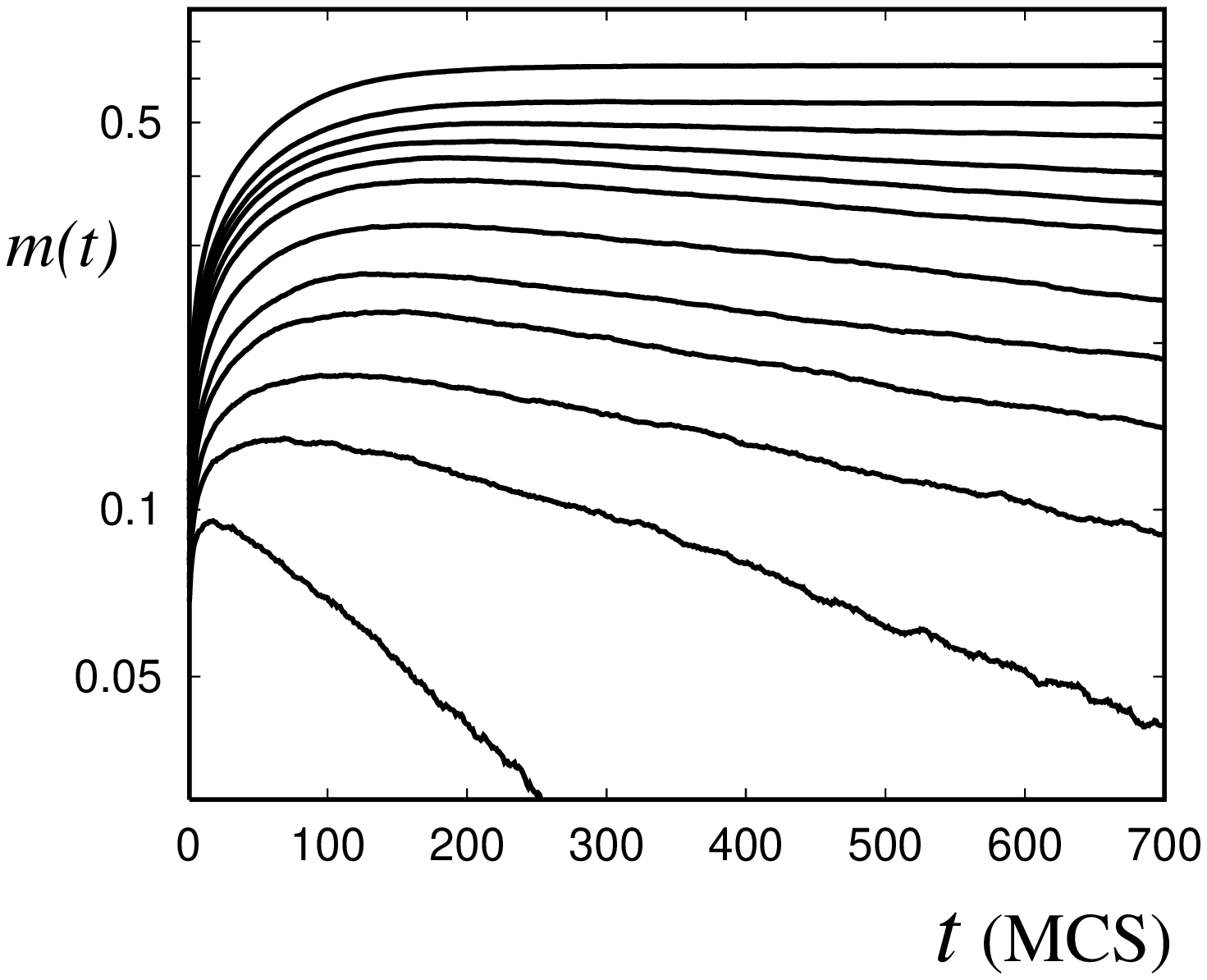}
\vspace*{-7.04cm}\\
\hspace*{9.4cm}
\noindent
\parbox{6cm}{{\small {\bf Fig 1.:} Magnetization
of the $6$-state clock model
as a function of time (in units of
Monte Carlo steps per spin) in semi-logarithmic representation
for $L\!=\!30$ and $m_0=0.05$
for various temperatures in, below, and above the KT
phase. From
bottom to top the corresponding temperatures are $T/J=1.1$,1.04, 1.0,
0.95, 0.92, 0.84, 0.76, 0.72, 0.68, 0.64, 0.6, 0.5. }}
\\[2.6cm]

\def\epsfsize#1#2{0.58#1}
\hspace*{-0.4cm}\epsfbox{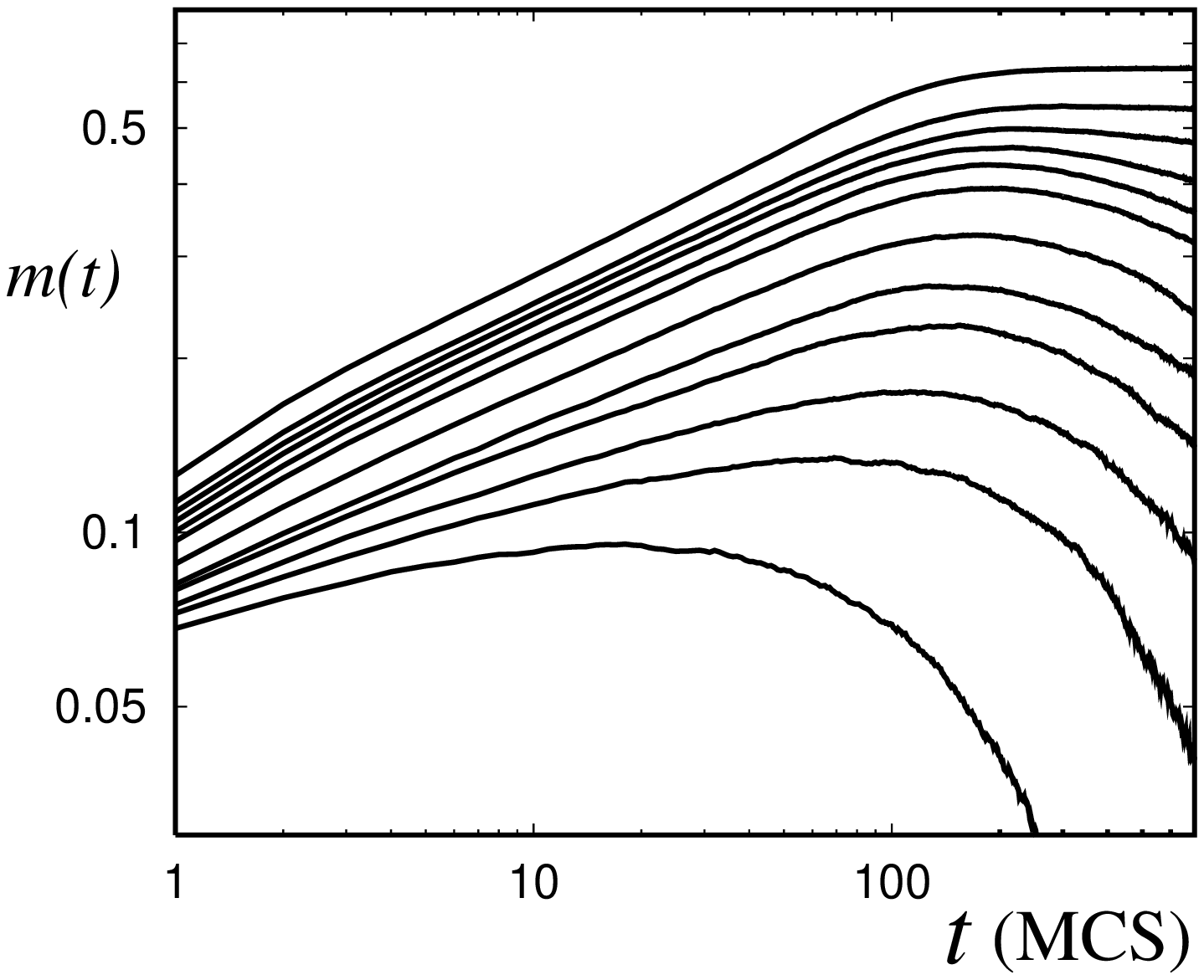}
\vspace*{-6.9cm}\\
\hspace*{9.4cm}
\noindent
\parbox{6cm}{{\small {\bf Fig 2.:} The same profiles as in Fig.\,1 in
double-logarithmic representation,
emphasizing the initial power-law growth of the magnetization.}}
\\[4.9cm]

The results are displayed in semi-logarithmic form in Fig.\,1. and
in double-logarithmic form in Fig.\,2. For
temperatures $T/J\lesssim 1.1$ the magnetization
has the tendency to increase initially. For the highest temperature
represented in the figure, this increase lasts only for about 10-20
updates of the lattice, but it becomes more and more pronounced
at lower temperatures. As one can see from Fig.\,2, a clearly identifiable
initial power-law regime at about $10\lesssim t\lesssim 100$
can be observed for $T/J \lesssim 0.95$. (The time is always
given in Monte Carlo steps per spin.)
All curves in the temperature range
$0.64\lesssim T/J\lesssim 1.1$ have a maximum
and thereafter decay exponentially or ``linearly''
$\sim \text{e}^{-t/t_L}$.

It is
instructive to extract the linear relaxation time
$t_L$ from the data in Fig.\,1. It depends on both the system
size and the temperature; results for $t_L(T)$ for $L=30$
are displayed in Fig.\,3 (squares
for $p=6$). The temperature
range of the KT phase as obtained by Challa and Landau \cite{chal}
is indicated in the figure. Obviously, in this quantity the signature
of the KT phase is a plateau-like structure,
roughly located between $T_1$ and $T_2$ of Ref.\,\cite{chal}.
Coming from the high-temperature side, $t_L$ increases
up to about $T/J=0.95$. Then $t_L$ remains almost unchanged
down to $T/J=0.75$ and starts to grow rapidly
when going to
lower temperatures. For $T/J\lesssim 0.6$ the magnetization
more or less approaches a constant within the observation time
$t<1200$, which means that $t_L$ becomes effectively infinite.
The latter is due to an exponentially growing tunneling time in the
ferromagnetic phase\cite{bind}.\\[1cm]

\def\epsfsize#1#2{0.58#1}
\hspace*{-0.4cm}\epsfbox{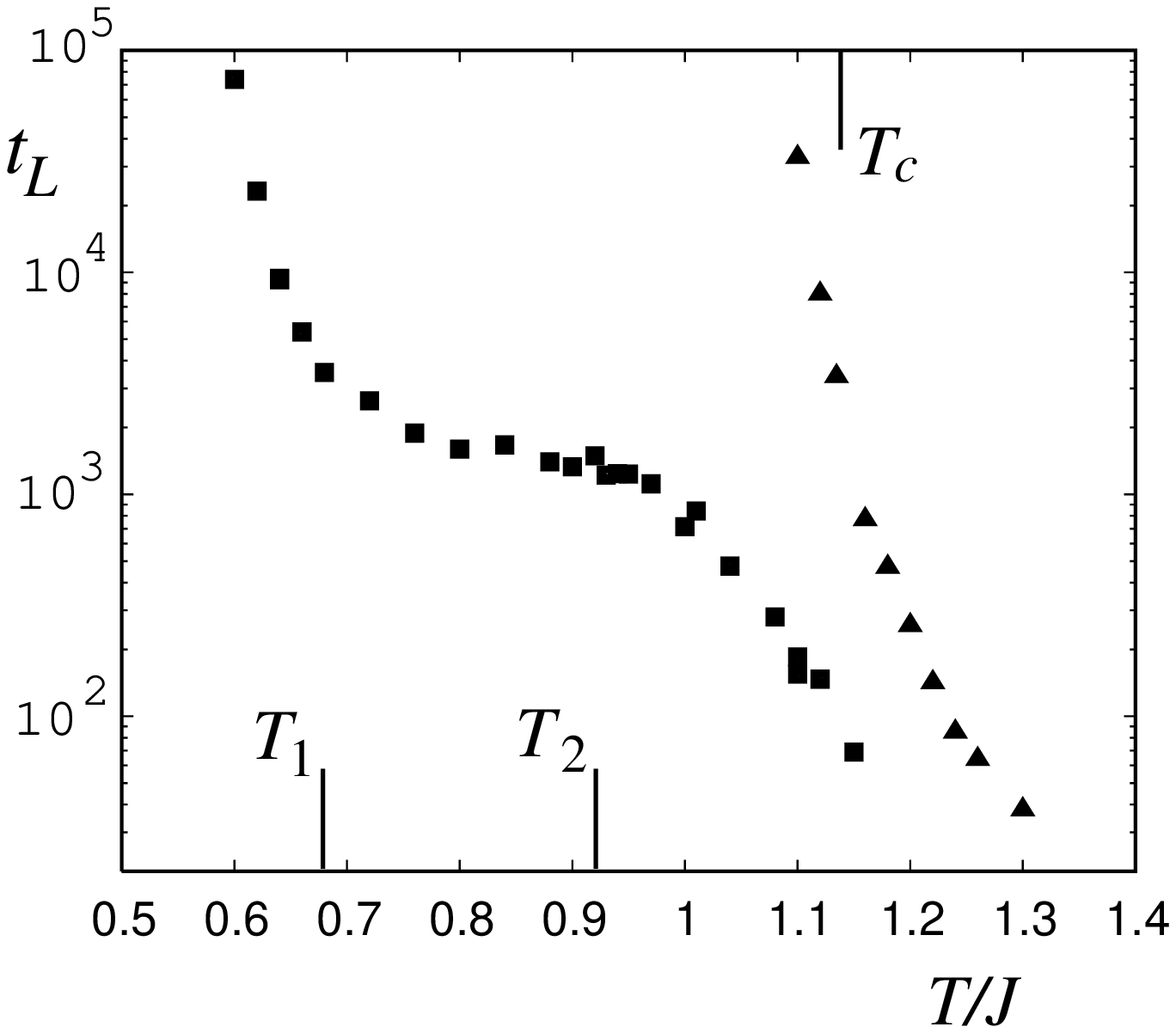}
\vspace*{-6.8cm}\\
\hspace*{9.4cm}
\noindent
\parbox{6cm}{{\small {\bf Fig 3.:} The linear relaxation times $t_L$
for $p=6$ (squares) and $p=4$ (circles)
as a function of
temperature for $L=30$, $m_0=0.05$. Both the (exact) critical temperature
of the 4-state clock model, $T_c/J=1.13..$, and the upper and
lower transition temperatures $T_1/J=0.68$ and $T_2/J=0.92$ determined
Challa and Landau\cite{chal} are indicated in the figure. The
error bars are approximately of the same size as the symbols.}}
\\[1cm]

For comparison we calculated a number of curves for the
4-state clock model in the vicinity of its (exact) critical
temperature $T_c/J=1.1345\ldots$. The corresponding linear relaxation
time is
also shown in Fig.\,3 (circles). It is steadily increasing
for decreasing temperature, as
one would have expected in the vicinity of a critical point, and no
plateau is seen.

In order to determine the short-time exponent $\theta$,
we calculated magnetization profiles for the lattice
size $L=300$. As also
found for other models\cite{sczh}, the outcome for
the short-time exponent
$\theta$ varies significantly with $m_0$.
{}From the renormalization-group
analysis it is clear \cite{jans,own} that the pure scaling form
(\ref{power}) is assumed for $m_0\to 0$
only. Unfortunately, it
is hard to obtain reliable results for $\theta$ in this limit,
as for $m_0=0$ the signal vanishes completely,
and for decreasing $m_0$ it becomes
more and more blotted out by the noise.
Hence, as in Ref.\,\cite{sczh}, we calculated profiles for a number
of initial values $m_0$ and then determined
$\theta$ by fitting
a power law to the short-time regime. As the fit interval we chose
$[10,300]$. The results for $\theta$ as a function of $m_0$
for the temperatures
$T/J=0.6$, 0.68, 0.76, 0.84, 0.92
are shown in Fig.\,5. The values of the
short-time exponent obtained by extrapolating linear fits to $m_0=0$
are listed in the Table.

Some representative
profiles for $L=300$ and $m_0=0.01$ are presented in Fig.\,4 (solid lines).
They show that in the given temperature range
the initial increase is consistent with a power law.
As seen
from the other curves displayed in Fig.\,4, this is
drastically different at low temperatures. E.g. for
$T/J=0.2$ (upper dashed curve), deep in the ferromagnetic regime,
the time dependence of the magnetization
is more complicated, and only for $t\gtrsim 200$ power-law behavior is assumed.
This is weakly indicated already at
$T/J=0.6$ (lower dashed line).\\[1cm]

\def\epsfsize#1#2{0.58#1}
\hspace*{-0.4cm}\epsfbox{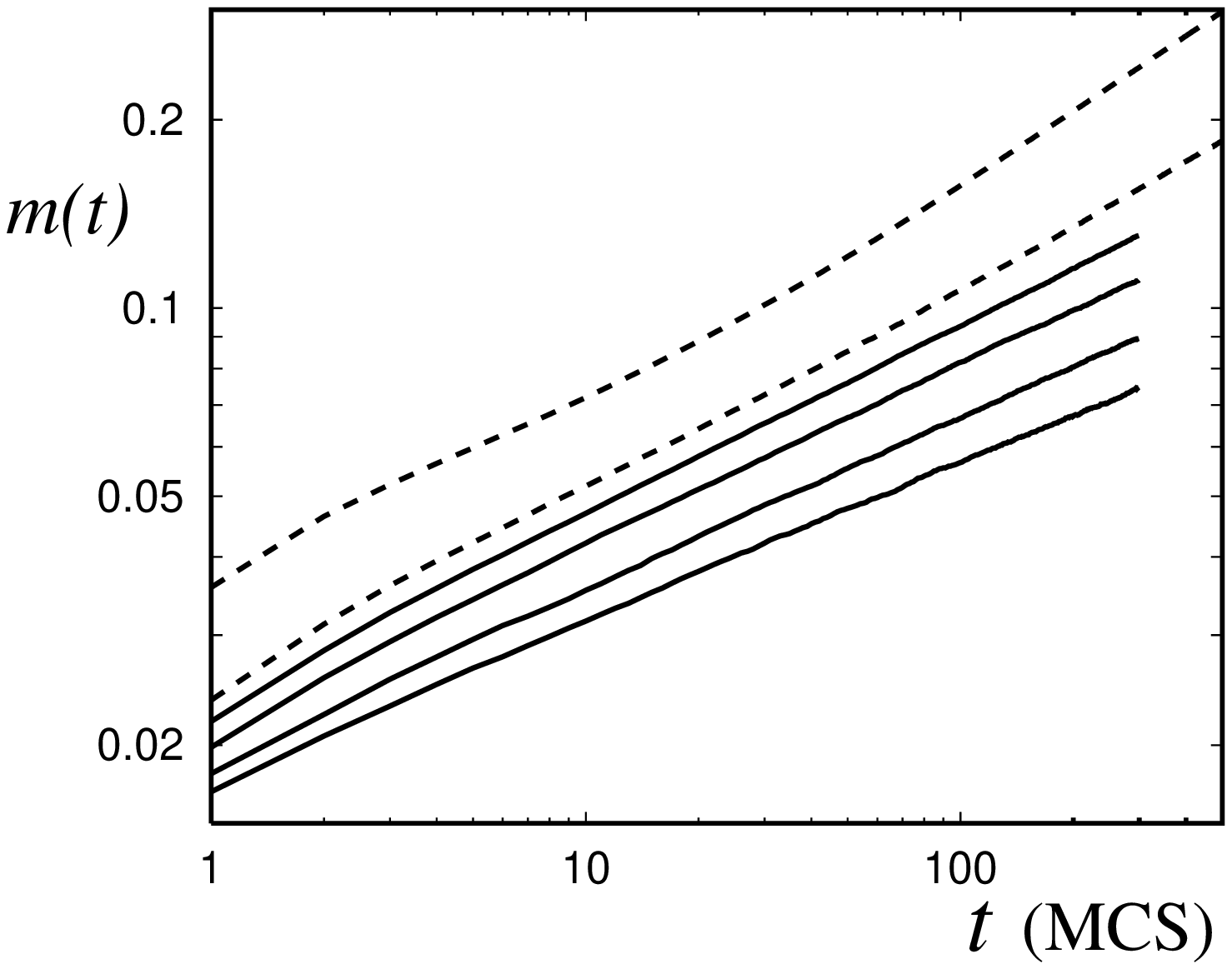}
\vspace*{-6.8cm}\\
\hspace*{9.4cm}
\noindent
\parbox{6cm}{{\small {\bf Fig 4.:} Initial behavior of the order parameter
for $L=300$
and $m_0=0.01$ for (from bottom to top) $T/J=0.92,\,0.84,\,
0.76,\,0.68,\,0.6,\,0.2$. The solid curves represent profiles
within the KT phase.}}
\\[5cm]

\def\epsfsize#1#2{0.58#1}
\hspace*{0cm}\epsfbox{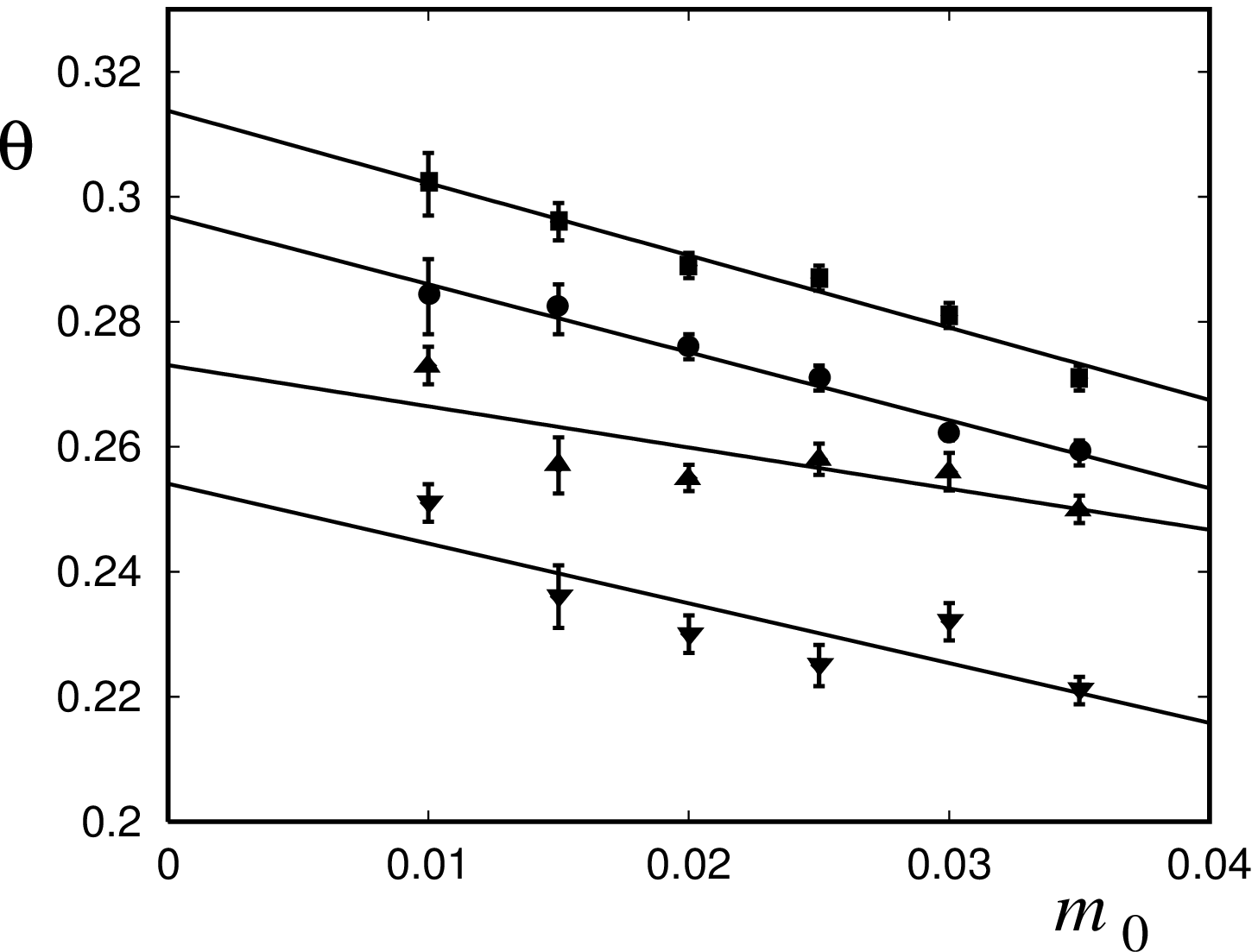}
\vspace*{-6.5cm}\\
\hspace*{9.4cm}
\noindent
\parbox{6cm}{{\small {\bf Fig 5.:} Results for the short-time exponent $\theta$
as
a function of $m_0$ for four temperatures $T/J=0.68$ (squares),
0.76 (full circles), 0.84 (triangles pointing up), and
0.92 (triangles pointing down). The respective
linear fits, from which
the extrapolations to $m_0=0$ were obtained, are also displayed.}}
\\[2.4cm]

\subsection{The autocorrelation function and the exponent $\lambda$}

The autocorrelation function $A(t)$ as defined in Eqn. (\ref{auto}) is
governed by the exponent $\lambda/z$.
Exemplary results for $A(t)$ in the KT phase
for $L=300$ (and $m_0=0$) are displayed in Fig.\,6
(solid lines). Even from this representation one can tell that
the curves are slightly curved up to $t\simeq 100-200$ and
that, if at all, they only relatively slowly approach a power law.
This is the reason why the exponent $\lambda/z$ in (\ref{lambda})
cannot be determined on smaller lattices with, say, $L=100$.
In this case $A(t)$ crosses over to the
finite-size (linear)
decay before a reliable estimate for $\lambda/z$ can be
extracted. For comparison we have also plotted the autocorrelation
function for the critical Ising model in Fig.\,6 (dashed curve).

In order to overcome this obstacle,
we calculated the autocorrelation
function for $L=300$ up to $t=1000$.
Then $\lambda/z$ was determined by fitting
the data in an interval $[t_1,t_2]$ within the
power-law regime.
To determine the minimal $t_1$
(called delay time in the following),
we calculated the difference
$\Delta A$ between the data and the fit for a given interval.
While the upper limit $t_2\simeq 1000$ was fixed, the lower limit $t_1$ was
increased until (besides random fluctuations) $\Delta A$ did
not show any systematic deviation from the fit
within $[t_1,t_2]$.
\\[0.5cm]

\def\epsfsize#1#2{0.58#1}
\hspace*{-0.4cm}\epsfbox{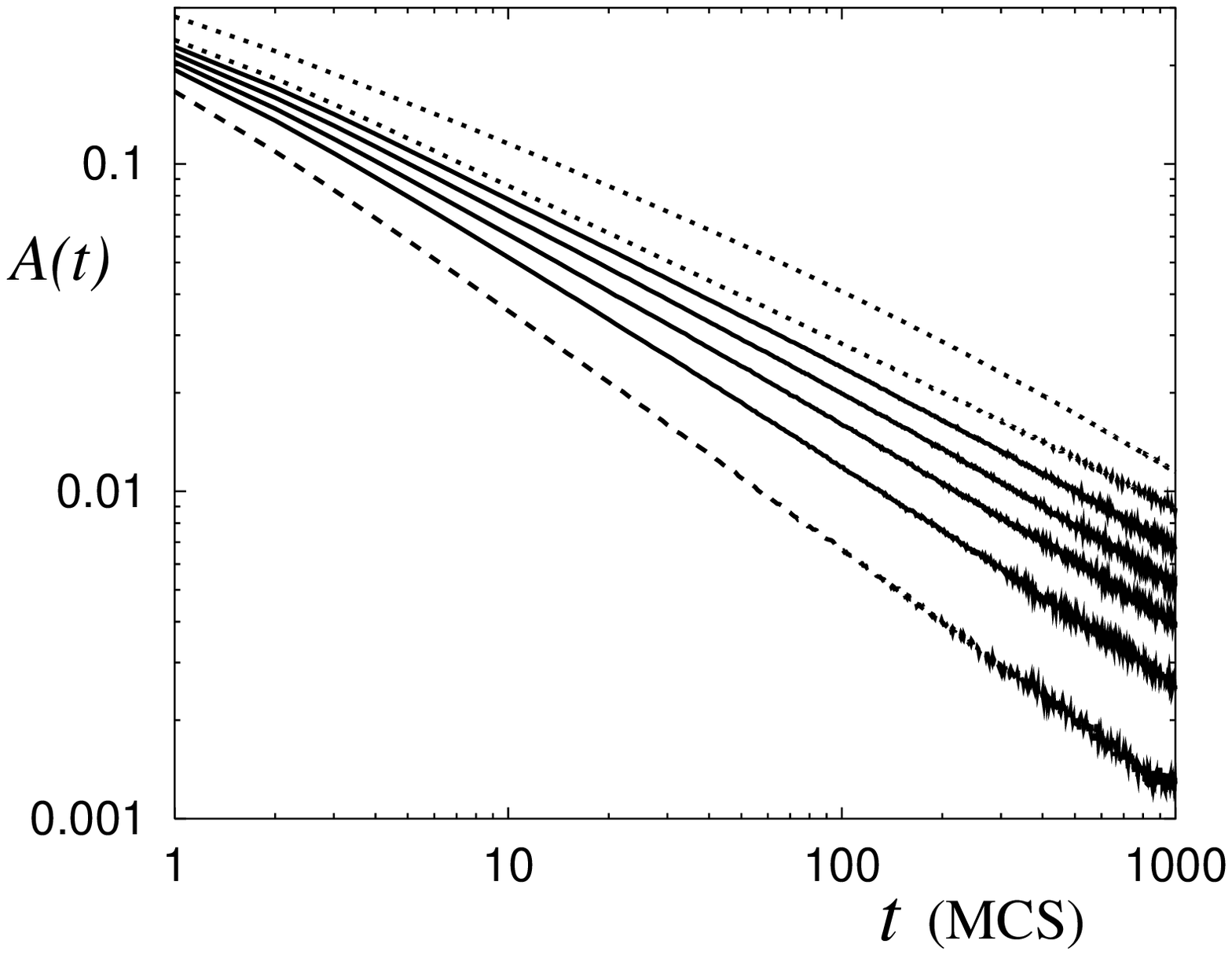}
\vspace*{-7.2cm}\\
\hspace*{9.4cm}
\noindent
\parbox{6cm}{{\small {\bf Fig 6.:} The autocorrelation function defined in
(\ref{auto})
for $L=300$, $m_0=0$ for $T/J=$0.92, 0.84,
0.76, 0.68, 0.6, 0.2 (upper six curves from bottom to top)
The dashed curve represents the data for the critical Ising
model.}}
\\[4.4cm]

Consider first the temperature $T/J=0.68$. Numerical results
for $\Delta A$ are shown
in Fig.\,7 (two upper curves). For the uppermost curve the fit
interval was $[200,1000]$. It is obvious that, besides the random
short-scale fluctuations, the data systematically deviate from the
fitted power law; for $200\lesssim t\lesssim 300$ the data are below
and for $400\lesssim t\lesssim 600$ they are systematically above zero.
In contrast, a fit to the interval $[500,1000]$ (second curve from above
in Fig.\,7) shows no systematic deviation, and, thus, for this temperature
the power-law regime seems to be reached for $t\simeq 500$.
\\[0.5cm]

\def\epsfsize#1#2{0.58#1}
\hspace*{-0.4cm}\epsfbox{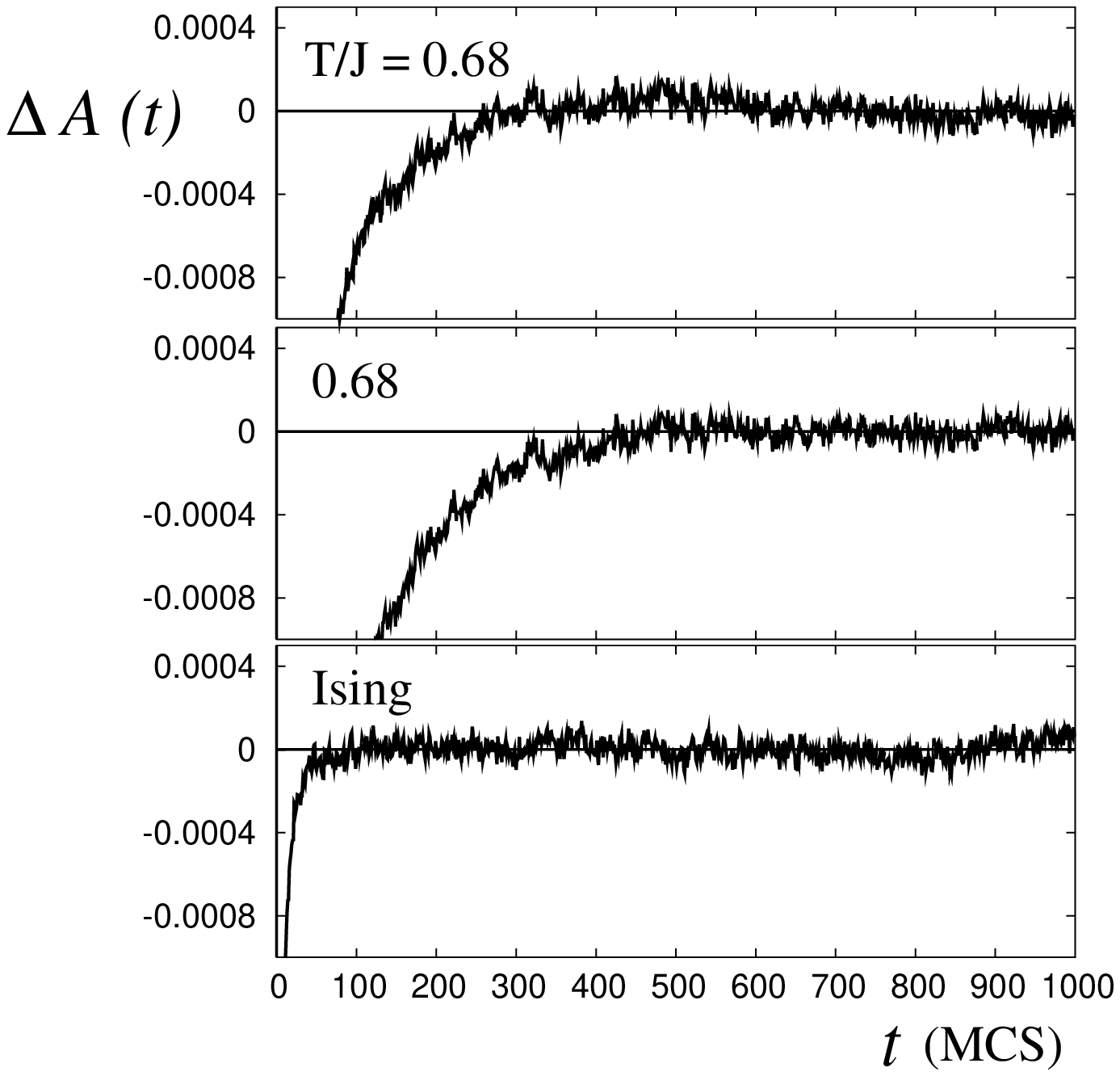}
\vspace*{-7.8cm}\\
\hspace*{9.4cm}
\noindent
\parbox{6cm}{{\small {\bf Fig 7.:} The difference $\Delta A(t)$
between the data for the autocorrelation
function and the power-law fit obtained for $T/J=0.68$
and fit intervals $[200,1000]$ (upper curve) and
$[500,1000]$ (middle curve). The bottom curve shows $\Delta A$ for
the critical Ising model and fit interval [80,1000].}}
\\[4.4cm]

In Fig.\,7 we have also plotted $\Delta A(t)$ for the critical
Ising model with fit interval $[80,1000]$. Although the delay time
$t_1\simeq 80$ is much shorter than for the clock model, it is
still relatively large compared with $m(t)$ or $m^{(2)}(t)$,
where one finds power-law behavior for $t\gtrsim 10$.
\\[0.5cm]

\def\epsfsize#1#2{0.58#1}
\hspace*{-0.4cm}\epsfbox{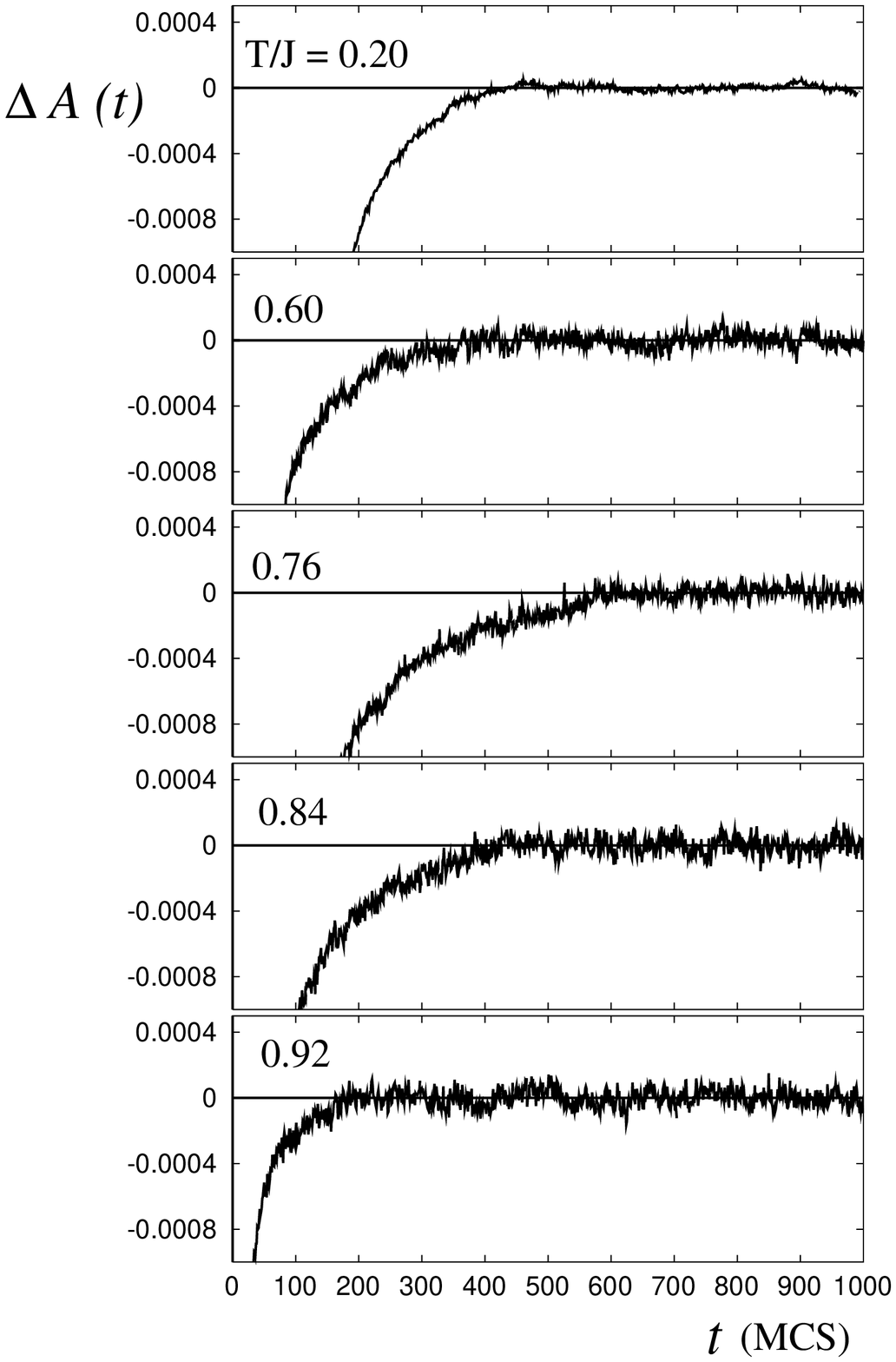}
\vspace*{-12.6cm}\\
\hspace*{9.4cm}
\noindent
\parbox{6cm}{{\small {\bf Fig 8.:} $\Delta A(t)$ for $T/J=0.92$,
0.84, 0.76, 0.6, and 0.2; the respective fit intervals are
[200,1000], [450,1000], [650,1000], [400,1000], and [500,1000].
 }}
\\[11cm]
In Fig.\, 8 the difference $\Delta A(t)$ is displayed for
various temperatures within and below the KT phase.
The estimated delay times are 500, 400, 500, 650, 450, and 200
for $T/J=0.2$, 0.6, 0.68, 0.76, 0.84, and 0.92, respectively.
Thus, $t_1$ is relatively short at $T_2$, increases
when $T$ is decreasing,
and appears to become smaller again for $T\lesssim T_1$.
However, presently our data are not accurate enough to
allow for definite statements about the temperature dependence
of $t_1$.

The results for $\lambda/z$ are shown in the Table.
The slow approach the power-law behavior
of $A(t)$ is a new feature
not seen in the Ising model. Presumably it is due to vortices
which are abundantly generated during the quench and slowly annihilate
during the ensuing relaxation.
The influence of vortices on the dynamics of the $XY$ model
was discussed by several authors in the literature.
Vortices in general give rise to new length scales\cite{Mondello,blun}, and,
as a consequence, the simple scaling picture discussed in Sect.\,II
breaks down. It was also shown that in certain quantities a
slow (logarithmic) approach to the
power-law behavior occurs\cite{Mondello,Yurke}, especially in those
which are substantially influenced by
short wavelengths fluctuations of the
order-parameter field.
However, with the help
of the existing literature it was not possible for us to get a complete
understanding of the observed dynamics at intermediate time scales.
Because we were mainly interested in the asymptotic power laws
and the exponents governing them, we
did not pursue this point any further.
\\[5mm]
\renewcommand{\arraystretch}{1.4}
\hspace*{1.4cm}
\begin{tabular}{||p{1.0cm}||p{1.7cm}
|p{1.7cm}|p{1.7cm}||p{1.7cm}|p{1.7cm}|p{1.7cm}||}\hline
$\>T/J$ & $\>\>\theta$ & $\>\>\lambda/z$ & $\>\>b$ & $\>\>z$ & $\>\>\eta$ &
$\>\>\eta_0-\eta$\\\hline\hline
0.92 & 0.254(5) & 0.670(10)& 0.816(12)&2.16(4)&0.23(6)&1.10(3)\\\hline
0.84 & 0.273(6) & 0.644(19)& 0.832(10)&2.18(6)&0.19(7)&1.19(4)\\\hline
0.76 & 0.297(2) & 0.622(11)& 0.842(12)&2.18(3)&0.17(5)&1.29(2)\\\hline
0.68 & 0.314(2) & 0.577(6)& 0.856(11)&2.24(2)&0.08(4)&1.41(2)\\\hline
0.60 & 0.355(5) & 0.517(6)& 0.838(11)&2.29(3)&0.08(5)&1.63(3)\\\hline
Ising & 0.191(2) & 0.730(6)& 0.806(2)&2.172(6)&0.25&0.83(1)\\\hline
\end{tabular}\\[4mm]

\noindent
{\small {\bf Table 1:} Monte Carlo estimates for the
exponents of the two-dimensional 6-state
clock model for temperatures within and slightly below
the KT phase. For comparison the
respective exponents
of the critical Ising model (at $T_c/J=2.2691\ldots$) are also displayed.
In the latter case $\theta$ and $z$ were taken from Ref.\,\cite{gras},
then $\lambda/z$, $b$, and $\eta_0-\eta$ were obtained with the help
of (\ref{lambda}), (\ref{b}), (\ref{theta}) using the exact $\eta=0.25$.
}\\[3mm]

\subsection{The second moment and results for $z$ and $\eta$}

The short-time behavior of the second moment $m^{(2)}(t)$ defined
in (\ref{mtwo}) is much less problematic than that of $A(t)$.
Some representative curves are shown
in Fig.\,9. As before, the results for the KT phase are
the solid curves.
The power-law behavior is found for $t\gtrsim 20$, and the exponent
$b$ in (\ref{b}) was determined in the interval $[20,500]$.
The results are listed in the Table.

Obviously, $b$
and because of (\ref{b}) the combination $(2-\eta)/z$ do not
depend much on the temperature. Taking into account
the errors, the outcome is even consistent with a temperature-independent
exponent, although a weak decrease for increasing temperature seems
also possible.

As can be further seen from Fig.\,9 also for
the critical Ising model (dashed curve)
this exponent is not much different. From our own data displayed in Fig\,9
we determined $b=0.81(1)$.
Taking alternatively the exact $\eta=0.25$ and the literature
value $z=2.172(6)$ \cite{gras}, one finds with the help of (\ref{b})
$b=0.806(2)$.
Finally, we use the scaling relation (\ref{lambda}) and (\ref{b})
to get estimates for the dynamic exponent $z(T)$,
the anomalous dimension $\eta(T)$, and
the difference $\eta_0-\eta$. The results are also displayed
in the Table.
\newpage
\vspace*{-.6cm}
\def\epsfsize#1#2{0.55#1}
\hspace*{-0.4cm}\epsfbox{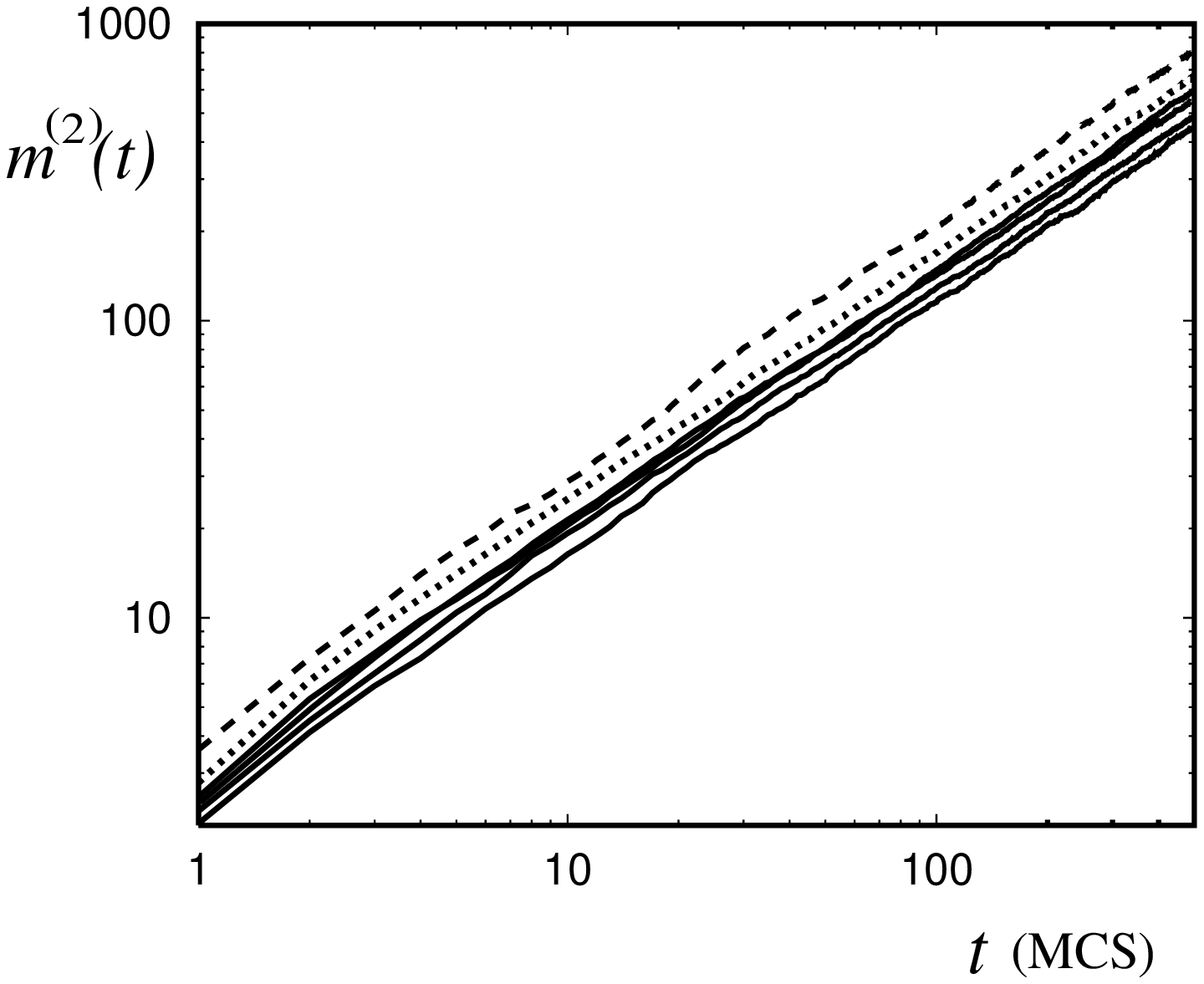}
\vspace*{-6.5cm}\\
\hspace*{9.4cm}
\noindent
\parbox{6cm}{{\small {\bf Fig 9.:} The second moment defined in (\ref{mtwo})
for $L=300$,
$m_0=0$ and $T=0.92,\,0.84,\,
0.76,\,0.68,\,0.6$ (lower five curves from bottom to top). The dashed
curve shows the data for the critical Ising model.}}
\\[3.1cm]

\section{Summary and Discussion}
We investigated the short-time relaxational behavior of the
two-dimensional clock model after a quench from
a high-temperature initial state to the KT phase.
Motivated by the situation in Ising-like systems, we assumed
that an independent initial exponent also exists for the clock
model, that it determines the short-time behavior in the KT
phase, and that the static exponent $\eta(T)$ can be
determined from the initial stage of relaxation processes.
Under the assumption
that the dynamics is asymptotically governed by only
one (macroscopic) length scale, the time-dependent
correlation length $\xi(t)\sim t^{1/z}$, we explored the consequences
of the anomalous short-time behavior.
Numerical results for the magnetization, its second moment and the
autocorrelation function were obtained by Monte Carlo simulation.

A new feature that is not found in the Ising model is a relatively slow
approach to a power law of the autocorrelation function. Presumably, this
is due to vortices and additional length scales associated with them.
{}From the existing literature we were not able to get a
full quantitative understanding of this phenomenon, in particular
in view of the relatively fast approach to the power law observed
in the magnetization and its second moment.

Another observation concerns the exponent $b$ of the second moment.
It turns out to be either weakly depending
on the temperature or even constant. It would be interesting to
study the temperature dependence of this quantity
also for other values of $p$ and for the $XY$ model ($p\to \infty$).

Our Monte Carlo
estimates for $\eta(T)$ listed in the Table are consistent
with earlier analytical and numerical work\cite{jose,chal}.
Since $\eta$ is a derived quantity, its error is relatively
large, and with the present procedure it would be hard to
obtain a much smaller value. However, we believe that these results
confirm our basic hypothesis
that the universal short-time behavior,
previously known from Ising-like
systems, also occurs in the KT phase.
In particular this implies that there exists a new exponent
$\eta_0(T)$, the
anomalous dimension of the
initial magnetization, that in general is different from the
exponent $\eta(T)$ of the static correlation function. This
difference gives rise to
the characteristic short-time behavior of the magnetization (\ref{power}) and
also leaves its fingerprint on the time dependence
of the autocorrelation function $A(t)$ as expressed in (\ref{lambda}).\\[2mm]
{\small {\bf Acknowledgements}: This work was supported in part by the
Deutsche Forschungsgemeinschaft through
Sonderforschungsbereich 237 ``Unordnung und gro{\ss}e Fluktuationen''.}


\begin{thebibliography} {99}
%
\bibitem{jans} H.\ K.\ Janssen, B.\ Schaub, and B.\ Schmittmann,
Z.\ Phys.\ B {\bf 73}, 539 (1989).
%
\bibitem{halp}
P.\ C.\ Hohenberg and B.\ I.\ Halperin,
Rev.\ Mod.\ Phys.\ {\bf 49}:435 (1977).
%
\bibitem{own} H.\ W.\ Diehl and U.\ Ritschel, J.\ Stat.\ Phys.
{\bf 73}, 1 (1993).
%
\bibitem{foo1} The short-time exponent $\theta$ was denoted by
$\theta'$
in Ref.\,\cite{jans}. But since the original $\theta$ of \cite{jans}
does not appear in the present work, we have dropped the prime.
%
\bibitem{monte} Z.-B. Li, U. Ritschel and B. Zheng, J. Phys. A:
Math. Gen.
{\bf 27}, L837 (1994).
%
\bibitem{gras} P. Grassberger, Physica A {\bf 214}, 547 (1995).
%
\bibitem{liea} Z.-B. Li, L. Sch\"ulke, B. Zheng, Phys. Rev. Lett.
{\bf 74}, 3396 (1995); {\it Finite-size scaling and critical
exponents in critical relaxation}, University of Siegen preprint
Si-95-04.
%
\bibitem{sczh}
L. Sch\"ulke and B. Zheng, {\it The Short-Time
Dynamics of the Critical Potts Model},
University of Siegen preprint Si-95-2.
%
\bibitem{jans2}{H.\ K.\ Janssen, in {\it From Phase Transitions
to Chaos --- Topics in Modern Statistical Physics},
edited by G.\ Gy\"orgyi, I.\ Kondor, L.\ Sasv\'ari, and T.\ Tel
(World Scientific, Singapore, 1992).}
%
\bibitem{oerding}{K. Oerding and H. K. Janssen, J. Phys. A:
Math. Gen. {\bf 26}, 3369 (1993);  {\it ibid.} {\bf 26}, 5295 (1993)}.
%
\bibitem{tricrit} H. K. Janssen
and K. Oerding, J. Phys. A:Math. Gen. {\bf 27}, 715 (1994).
%
\bibitem{koth} J. M. Kosterlitz and D. J. Thouless, J. Phys. C {\bf 6}, 1181
(1973); J. M. Kosterlitz, J. Phys. C {\bf 7}, 1046 (1974).
%
\bibitem{jose} V. Jos\'e, L. P. Kadanoff, S. Kirkpatrick, and
D. R. Nelson, Phys. Rev. B {\bf 16}, 1217 (1977); see also
D. A. Nelson, in {\it Phase Transitions and
Critical Phenomena}, Eds.: Domb and J. L. Lebowitz,
(London, Academic Press, 1983)
%
\bibitem{mewa} N. D. Mermin and H. Wagner, Phys. Rev. Lett. {\bf 22},
1133 (1966).
%
\bibitem{p6cl} H. V. Roomany and H. W. Wyld,
Phys. Rev. B {\bf 23}, 1357 (1981);
P. Rujan, G. O. Williams, and H. L. Frisch,
Phys. Rev. B {\bf 23}, 1362 (1981).
%
\bibitem{tobo} J. Tobochnik, Phys. Rev. B {\bf 26}, 6201 (1982);
{\it ibid.} {\bf 27}, 6972 (1983).
%
\bibitem{kagu} K. Kaski and J. D. Gunton,
Phys. Rev. B {\bf 28}, 5371 (1983);
K. Kaski, M. Grant, and J. D. Gunton,
Phys. Rev. B {\bf 31}, 3040 (1985).
%
\bibitem{chal} M. S. S. Challa and D. P. Landau,
Phys. Rev. B {\bf 33}, 437 (1986).
%
\bibitem{yaon} A. Yamagata and I. Ono, J. Phys. A
%
\bibitem{lero} Y. Leroyer and K. Rouditi, J. Phys. A {\bf 24}, 1931 (1991).

%
\bibitem{Mondello}
M. Mondello and N. Goldenfeld, Phys. Rev. A {\bf 42}, 5865 (1990).
%
\bibitem{Yurke} B. Yurke, A. N. Pargellis, T. Kovacs, D. A. Huse,
Phys. Rev. E {\bf 47}, 1525 (1993).
%
\bibitem{blun} R. E. Blundell and A. J. Bray, Phys. Rev. E {\bf 49},
4925 (1994); A. J. Bray and A. D. Rutenberg, Phys. Rev. E {\bf 49},
R27 (1994).
%
\bibitem{foo3} Throughout this paper we use the ``magnetic''
language. The term order parameter is sometimes used
as a synonym for magnetization (density) and does not mean
any form of topological order, which would
characterize the difference between
{\it equilibrium} states above and below
$T_{\text{KT}}$.
%
\bibitem{foo5} Since we identify $\eta/2$ with
the scaling dimension of the field,
all formulas presented in Sect.\,III are
restricted to $d=2$. But it
is straightforward to extent the analysis to general spatial
dimension.
%
\bibitem{huse} D. A. Huse, Phys. Rev. B {\bf 40}, 304 (1989).
%
\bibitem{own3} U. Ritschel and H. W. Diehl, {\it Dynamical
relaxation and universal short-time behavior in finite systems:
the renormalization-group approach}, University of Essen preprint.
%
\bibitem{own2} U. Ritschel and H. W. Diehl, Phys. Rev. E {\bf 51}, 5392
(1995).
%
\bibitem{sancho} J. M. Sancho, M. San Miguel, and J. D. Gunton,
J. Phys. A {\bf 13}, L443 (1980).
%
\bibitem{bind} K. Binder and D. W. Heermann, {\it Monte Carlo Simulation
in Statistical Physics}, (Springer Verlag, Berlin, 1988).
%
\end{thebibliography}
\end{document}